\documentclass[twocolumn,superscriptaddress,prb]{revtex4}

\usepackage{graphicx}

\begin{document}

\title{Onset of optical-phonon cooling in multilayer graphene revealed by RF noise and black-body radiation thermometries}

\author{D. Brunel}
\affiliation{Laboratoire Pierre Aigrain, Ecole Normale Sup\'erieure-PSL Research University,
CNRS, Universit\'e Pierre et Marie Curie-Sorbonne Universit\'es,
Universit\'e Paris Diderot-Sorbonne Paris Cit\'e, 24 rue Lhomond, 75231 Paris Cedex 05, France}

\author{S. Berthou}
\affiliation{Laboratoire Pierre Aigrain, Ecole Normale Sup\'erieure-PSL Research University,
CNRS, Universit\'e Pierre et Marie Curie-Sorbonne Universit\'es,
Universit\'e Paris Diderot-Sorbonne Paris Cit\'e, 24 rue Lhomond, 75231 Paris Cedex 05, France}

\author{R. Parret}
\affiliation{Laboratoire Pierre Aigrain, Ecole Normale Sup\'erieure-PSL Research University,
CNRS, Universit\'e Pierre et Marie Curie-Sorbonne Universit\'es,
Universit\'e Paris Diderot-Sorbonne Paris Cit\'e, 24 rue Lhomond, 75231 Paris Cedex 05, France}

\author{F. Vialla}
\affiliation{Laboratoire Pierre Aigrain, Ecole Normale Sup\'erieure-PSL Research University,
CNRS, Universit\'e Pierre et Marie Curie-Sorbonne Universit\'es,
Universit\'e Paris Diderot-Sorbonne Paris Cit\'e, 24 rue Lhomond, 75231 Paris Cedex 05, France}

\author{P. Morfin}
\affiliation{Laboratoire Pierre Aigrain, Ecole Normale Sup\'erieure-PSL Research University,
CNRS, Universit\'e Pierre et Marie Curie-Sorbonne Universit\'es,
Universit\'e Paris Diderot-Sorbonne Paris Cit\'e, 24 rue Lhomond, 75231 Paris Cedex 05, France}

\author{Q. Wilmart}
\affiliation{Laboratoire Pierre Aigrain, Ecole Normale Sup\'erieure-PSL Research University,
CNRS, Universit\'e Pierre et Marie Curie-Sorbonne Universit\'es,
Universit\'e Paris Diderot-Sorbonne Paris Cit\'e, 24 rue Lhomond, 75231 Paris Cedex 05, France}

\author{G. F\`eve}
\affiliation{Laboratoire Pierre Aigrain, Ecole Normale Sup\'erieure-PSL Research University,
CNRS, Universit\'e Pierre et Marie Curie-Sorbonne Universit\'es,
Universit\'e Paris Diderot-Sorbonne Paris Cit\'e, 24 rue Lhomond, 75231 Paris Cedex 05, France}

\author{J.-M. Berroir}
\affiliation{Laboratoire Pierre Aigrain, Ecole Normale Sup\'erieure-PSL Research University,
CNRS, Universit\'e Pierre et Marie Curie-Sorbonne Universit\'es,
Universit\'e Paris Diderot-Sorbonne Paris Cit\'e, 24 rue Lhomond, 75231 Paris Cedex 05, France}

\author{P. Roussignol}
\affiliation{Laboratoire Pierre Aigrain, Ecole Normale Sup\'erieure-PSL Research University,
CNRS, Universit\'e Pierre et Marie Curie-Sorbonne Universit\'es,
Universit\'e Paris Diderot-Sorbonne Paris Cit\'e, 24 rue Lhomond, 75231 Paris Cedex 05, France}

\author{C. Voisin}
\email{christophe.voisin@lpa.ens.fr}
\affiliation{Laboratoire Pierre Aigrain, Ecole Normale Sup\'erieure-PSL Research University,
CNRS, Universit\'e Pierre et Marie Curie-Sorbonne Universit\'es,
Universit\'e Paris Diderot-Sorbonne Paris Cit\'e, 24 rue Lhomond, 75231 Paris Cedex 05, France}

\author{B. Pla\c{c}ais}
\email{bernard.placais@lpa.ens.fr}
\affiliation{Laboratoire Pierre Aigrain, Ecole Normale Sup\'erieure-PSL Research University,
CNRS, Universit\'e Pierre et Marie Curie-Sorbonne Universit\'es,
Universit\'e Paris Diderot-Sorbonne Paris Cit\'e, 24 rue Lhomond, 75231 Paris Cedex 05, France}

\begin{abstract}

We report on electron cooling power measurements in few-layer graphene excited by Joule heating by means of a new setup combining electrical and optical probes of the electron and phonon baths temperatures. At low bias, noise thermometry allows us to retrieve the well known acoustic phonon cooling regimes below and above the Bloch Gr\"uneisen temperature, with additional control over the phonon bath temperature. At high electrical bias, we show the relevance of direct optical investigation of the electronic temperature by means of black-body radiation measurements that provide higher accuracy than noise thermometry. In this regime, the onset of new efficient relaxation pathways involving optical modes is observed.

\end{abstract}

\maketitle

\begin{figure}
\includegraphics[width=7cm]{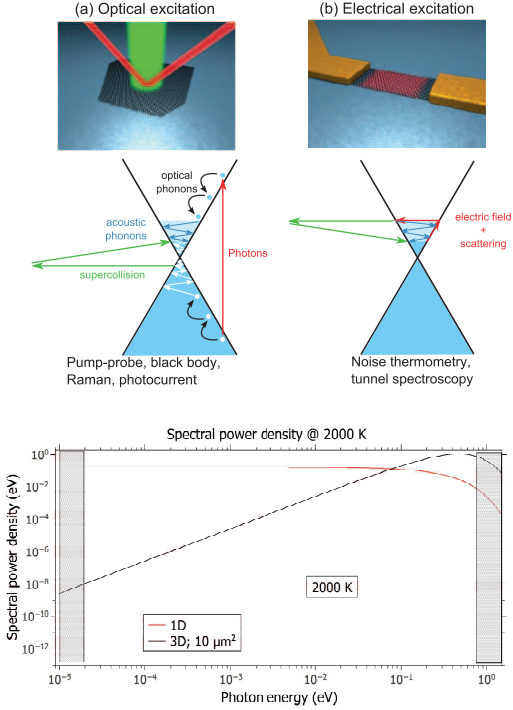}
\caption{Upper : Sketch of various electronic relaxation processes in monolayer graphene. Panel-a : photoexcited electron hole pairs relax by optical phonon emission (black trace) at high energy and acoustic  phonon emission at low energy (blue/white traces). Panel-b: electrical Joule heating produces a well controlled hot electron or hole population that also relaxes by acoustic phonon emission (blue traces). Impurity phonon scattering processes (supercollisions) are sketched in green. Lower : power density spectrum of a photon bath at $2000\;\mathrm{K}$ radiating in a 1D electromagnetic medium such as the Johnson Nyquist noise emitted by the hot electron bath of a matching load in a coaxial line and similar power density for a 10 $\mu$m$^2$ large conventional black-body radiating in an open 3D space. The dashed regions correspond to the instrumentally accessible regimes for RF and black-body detection schemes. } \label{Brunelfig1}
\end{figure}

Electron-phonon physics is at the crossroads of optics and electronics. Phonons control the relaxation of  photo-excited electron-hole pairs as well as that of electrons or holes when they are accelerated at high field. The confrontation of the two approaches is especially relevant in graphene where acoustic and optical phonons (ACs and OPs respectively)  have very contrasted effects, inherited from crystal symmetry and enforced by a strong carbon sp2-bonding.  OP-coupling is strong but restricted to high energies leaving a broad low-energy window  $\epsilon\lesssim 160\;\mathrm{meV}$ for the study of acoustic phonon effects. AC-phonon coupling is weak, due to large Fermi and phonon velocities ($v_F\simeq10^6\;\mathrm{m/s}$ and $s\simeq2.1\times 10^4\;\mathrm{m/s}$ respectively); it gives rise to prominent hot-carrier effects. The contrast between OPs and ACs effects is such that premises of OP cooling are expected for electronic temperatures below the OP phonon energy by relaxation of the most energetic electrons.
Hot electron populations can be generated and observed by both optical and transport approaches as sketched in Fig.~\ref{Brunelfig1}. In optics  a controlled light absorption (2.3 percent per graphene layer) generates high-energy ($\epsilon\sim1\;\mathrm{eV}$) electron and hole populations that relax into a hot electron or hole populations via OP-phonon emission and/or electron-electron interactions (Fig.~\ref{Brunelfig1}-a). The latter encompasses carrier multiplication and Auger scattering  \cite{Tielrooij2013nphys,Winzer2013prb}. As it is not relevant for the energy window investigated in the present study, we have omitted it in the figure for clarity. In the electrical scheme electrons (or holes) are drifting in an applied electric field until they get back-scattered by impurities and thermalize by electron-electron interactions (Fig.\ref{Brunelfig1}-b). In both cases one is left with a hot carrier distribution, at a temperature  $T_e\lesssim 2000\;\mathrm{K}$ ($\epsilon\lesssim 160\;\mathrm{meV}$), controlled by the balance between  light absorption or the  Joule power and AC-phonon emission.
Emission can be direct or assisted by impurity scattering; the later, called supercollision,  allows for larger phonon recoil energy.  Hot carrier detection can be achieved by photocurrent \cite{Graham2013nphys}, pump-probe technics \cite{Graham2014nl}, black-body radiation \cite{Freitag2010nnanotech, Berciaud2010} or noise thermometry \cite{Betz2012prl,Fong2012prx}.

In order to gain a deep insight into these relaxation processes, it is important to have a comprehensive diagnosis of the relevant electron, phonon and possibly photon temperatures. To this end, a versatile setup (hereafter called Janus setup) was developed that provides two primary electron temperature measurements through microwave noise and black-body radiation (see figure~\ref{Brunelfig1}) and several phonon temperature measurements schemes through Raman spectroscopy using either the direct Stokes/anti-Stokes ratio to monitor the occupation number of specific modes \cite{Steiner2009, Chae2010, Berciaud2010} or the anharmonic shift and broadening of optical modes as a probe of the AC mode reservoir \cite{Lang1999, Calizo2007}. In addition, this setup allows us to independently tune the phonon base temperature. In this introductory paper, we test and calibrate the performances of this new setup through the study of electron - AC phonon coupling in a few layer graphene sample and by comparing the data with the well established results of ML graphene. We further push the excitation (either electrical or optical) to higher values and open the way to the investigation of the coupling of electrons to optical phonons or substrate remote surface modes.

\begin{figure}[hh]
\includegraphics[width=8cm]{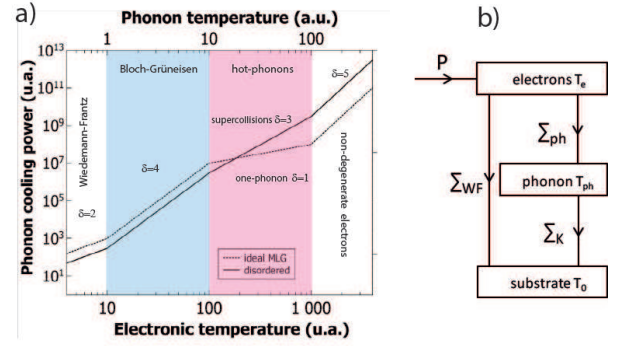}
\caption{Panel a) : different cooling pathways for electrons in graphene for increasing electron and acoustic phonon temperatures. They are characterized by the exponent $\delta$  of the cooling power $P$ as function of the electronic temperature $T_e$ (see details in the text). The coupling coefficients:  $\Sigma_{WF}$ for direct electron conduction to the contact at a temperature $T_0\sim4\;\mathrm{K}$, $\Sigma_{ph}$ for electron-phonon and $\Sigma_{K}$ for phonon substrates are sketched in panel b).
}\label{Brunelfig2}
\end{figure}

Before getting into the details of our experiment, we review below the basics of AC-phonon cooling in monolayer graphene (MLG), which is a well established framework  useful for further studies. The coupling of electrons to AC phonons has been extensively investigated theoretically \cite{Kubakaddi2009prb,Tse2009prb,Bistritzer2009prl,Viljas2010prb,Song2012prl,Chen2012prb,Virtanen2014prb} and experimentally mostly in monolayer graphene  \cite{Betz2012prl,Fong2012prx,Betz2013nphys,Laitinen2014nl} but also in bilayer graphene (BLG) \cite{Fay2011prb,Laitinen2014arXiv}. The heat sink by AC-phonon is characterized by a power law $P=\Sigma(T_e^\delta-T_{ph}^\delta)$, where $T_e$ and $T_{ph}$ are the electronic and AC-phonon temperatures, $\Sigma$ is the coupling constant and $\delta$ is an exponent that is characteristic of the cooling regimes.  For supported graphene at low substrate temperatures $T_0$ the phonon bath is  cooled according to the Kapitza law  $P\simeq\Sigma_K(T_{ph}^4-T_{0}^4)$ where $\Sigma_K\sim 1$--
$10\;\mathrm{Wm^{-2}K^{-4}}$. For typical  powers $P\sim 10^4$--$10^8\;\mathrm{Wm^{-2}}$ one can thus achieve $T_{ph}\sim10$--$100\;\mathrm{K}$. In these conditions different regimes can be observed that are sketched in Fig.\ref{Brunelfig2} for clean MLGs  following Ref.\cite{Viljas2010prb}. In particular the phonon temperature range covers the crossover from the cold-phonon (or Bloch-Gr\"uneisen) regime where $\delta=4$ to the hot-phonon (or equipartition) regime where $\delta=1$ (dashed line for clean graphene in Fig.\ref{Brunelfig2}). The crossover temperature $T_{BG}$ corresponds to the situation where the thermal phonon momentum reaches twice the Fermi momentum; $T_{BG}=2s/v_F\times T_F$ is thus proportional to the Fermi temperature $T_F=\epsilon_F/k_B$ and falls in the range $T_{BG}\sim50\;\mathrm{K}$ ($k_BT_{BG}\gtrsim 4\;\mathrm{meV}$) in doped graphene ($n\sim10^{12}\;\mathrm{cm^{-2}}$) where $T_F\sim1000\;\mathrm{K}$ ($\epsilon_F\sim 80\;\mathrm{meV}$). At very large electronic temperatures (or low $\epsilon_F$) the electron gas eventually  becomes non degenerate, a regime where $\delta=5$. At the low temperature end, the direct heat conduction to the leads takes over and $P=(4k_BT_e/e)^2/(3RLW)$ is given by the  Wiedemann Franz law where $R$, $L$ and $W$ are the sample resistance, length and width. The picture emerging from experiments is reminiscent of the above scenario for ideal graphene with however a qualitative difference which is the supercollision regime with $\delta=3$ overcoming the one-phonon regime with $\delta=1$ for $T_{ph}>T_{BG}$ \cite{Song2012prl,Graham2013nphys,Betz2013nphys,Laitinen2014nl} in the presence of disorder or multi-phonon processes \cite{Song2012prl,Virtanen2014prb}. Note that disorder also affects the value of $\Sigma$ but not of the exponent for $T_{ph}<T_{BG}$ \cite{Betz2012prl}. Transitions where reported from Wiedemann-Franz to Bloch-Gr\"uneisen regimes ($T^2\rightarrow T^4$ in Ref.\cite{Betz2012prl}), from Wiedemann-Franz to supercollisions ($T^4\rightarrow T^3$ in Ref.\cite{Betz2013nphys}) from non-degenerate to degenerate ($T^3\rightarrow T^5$ in Ref.\cite{Laitinen2014nl}). To our knowledge the one hot phonon regime with $P\propto T$, expected in ideal graphene, has never been reported.  This and other questions concerning the case of multilayer graphene, and the extreme electronic temperatures when optical phonons come into play, remain to be addressed. It is the main motivation for the present study.

\begin{figure}[hh]
\includegraphics[width=9cm]{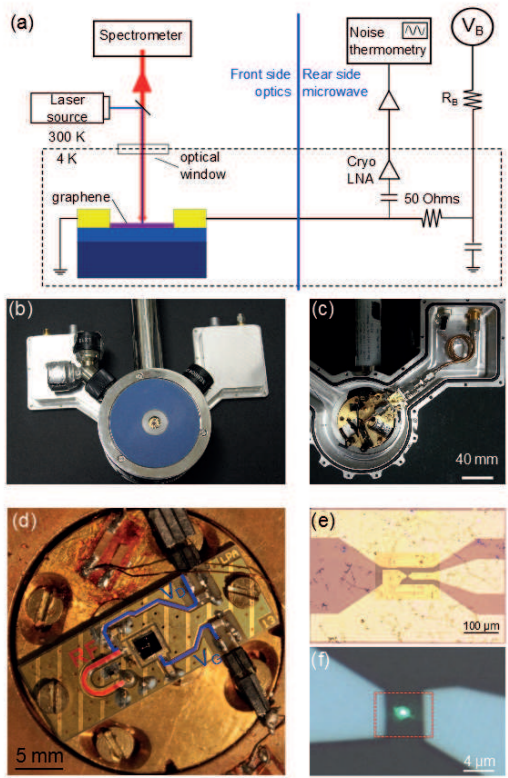}
\caption{The Janus set-up designed for opto-electronic characterization of graphene. The graphene sample is exposed to full optical investigation in the front and connected to the RF detection in the back side by an RF coaxial via-line. A detailed description of the panels is given in the text.}\label{Brunelfig3}
\end{figure}

The Janus setup is presented in Figure \ref{Brunelfig3}. It has two faces : a front side which is used for optical spectroscopy (Fig.\ref{Brunelfig3}-b) and a back side used for high resolution microwave measurement including noise thermometry (Fig.\ref{Brunelfig3}-c). The working principle is sketched in Fig. \ref{Brunelfig3}-a. The sample is located at the front, wire bonded to a printed circuit board PCB (Fig. \ref{Brunelfig3}-d) connected to the RF electrical components by an RF line in the center of the cold plate connecting the two sides. Janus is based on a variable temperature optical cryostat customized with an aluminium extension (Fig. \ref{Brunelfig3}-b) designed to host a miniature cryogenic, $0.1$--$2\;\mathrm{GHz}$ bandwidth, low noise amplifier. A $650\;\mathrm{mm}$-long Cu-Be coaxial cable is used to connect the LNA to the room temperature stage, while ensuring thermal decoupling.  Figure \ref{Brunelfig3}-d shows the PCB, where the biasing and RF lines have been
colored in blue and red respectively while the ground plane is signaled by zebra. The PCB is plugged in an MMCX connector in the back (not-shown).
The graphene sample (Fig.\ref{Brunelfig3}-f) is embedded in an on-chip coplanar waveguide (Fig.\ref{Brunelfig3}-e) for electrical measurement and illuminated by a $532\;\mathrm{nm}$-laser spot used for photo excitation and/or Raman spectroscopy. Experiments presented here were performed at the base temperature $T_0\simeq5\;\mathrm{K}$.

\begin{figure}[hh]
\includegraphics[width=7cm]{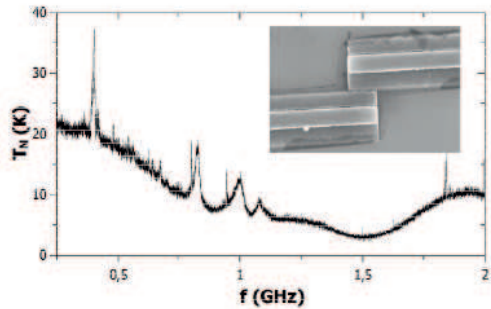}
\caption{Noise floor of the Janus setup. Main panel shows the spectrum of the noise temperature of the setup; it is calibrated by a tunnel junction  (SEM picture in the inset) as detailed in the text. }\label{Brunelfig4}
\end{figure}

 In the noise thermometry scheme, the time dependant voltage signal $V(t)$ is measured by a fast numerical oscilloscope with a $5\;\mathrm{Gsample/s}$ sampling rate and a $25 000\;\mathrm{sample}$ record length. The current noise $\delta I(t)$ generated by a sample, of typical impedance $R\sim 1\;\mathrm{kOhm}$, is converted in a voltage fluctuation at the $50\;\mathrm{Ohm}$ matching load and amplified by a RF line of overall power gain $G\simeq 85\;\mathrm{dB}$. The first stage is a cryogenic low noise amplifier, fitted in the Janus setup as shown in Figure\ref{Brunelfig3}-c. It controls the  $0.1$--$2\;\mathrm{GHz}$ measuring bandwidth and the noise resolution. Spectra are calculated up to $2.5\;\mathrm{GHz}$ with a $153\;\mathrm{kHz}$ resolution. For calibration purpose, the sample is substituted by an Al/AlOx/Al tunnel junction of similar impedance (inset of Figure \ref{Brunelfig4}) which acts as a primary noise standard providing a white current noise  $S_I = 2eI$ over a bandwidth of $\gtrsim 5\;\mathrm{
GHz}$ limited by the junction capacitance.
Calibration is carried out as follows:  $S_{I}(f)$ spectra of the output noise are recorded in the range $f=0.1-2.5\;\mathrm{GHz}$ as function of $I$ and fitted to a linear  current dependence to deduce both $G(f)$  and  the total noise of the amplification line, expressed in terms of input noise temperature $T_N(f)\equiv50S_I(f)/4k_B$. The later incorporates the bath temperature, $T_0\sim 5\;\mathrm{K}$ and the excess noise of the cryogenic LNA. As seen in figure \ref{Brunelfig4}, the noise temperature background remains below $10\;\mathrm{K}$ over a $1\;\mathrm{GHz}$ bandwidth giving a $1\;\mathrm{K}$ noise temperature resolution that translates into a $20\;\mathrm{K}$ resolution for the graphene sample electronic temperature. The resolution is quite similar to that achieved in a conventional helium bath cryostat (see Ref.\cite{Betz2012thesis}). The spectrum $T_N(f)$ being stable from run to run is used as a subsidiary noise standard in the data reduction.

\begin{figure}[hh]
\includegraphics[width=7cm]{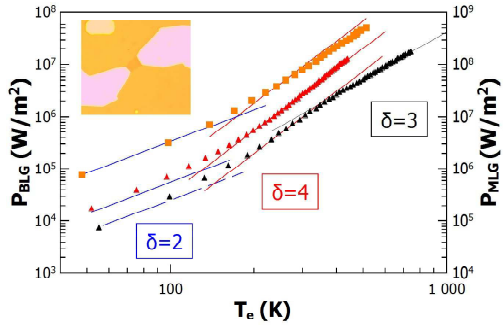}
\caption{Acoustic phonon cooling (orange squares) in a naturally doped bilayer graphene (BLG) deposited on silica. The sample (SEM picture in the inset) has dimensions $L\times W\simeq5\times3.3\;\mathrm{\mu m^2}$. The $P(T_e)$ dependence is very similar to previous measurements (ref.\cite{Betz2013nphys}) in p-doped monolayer graphene (MLG) with densities $n=4$ and $8\times 10^{11}\;\mathrm{cm^{-2}}$ (black and red triangles, axis on the right). Solid lines are power laws $P\propto T_e^\delta$ illustrating three cooling regimes sketched in Fig.\ref{Brunelfig2} : electron conduction ($\delta=2$ blue lines), Bloch-Gr\"uneisen one-phonon regime ($\delta=4$ red lines) and electron-phonon-impurity supercollisions ($\delta=3$ black line) which are observed in increasing electron temperatures.  }\label{Brunelfig5}
\end{figure}

As a first test of the Janus setup we have reproduced a standard acoustic phonon cooling experiment on a bilayer graphene sample  (figure \ref{Brunelfig5}). The sample is exfoliated on SiO$_2$; its dimensions are  $L\times W=5\times3.3\;\mu\mathrm{m}$ and its resistance was $R=1350\;\mathrm{Ohms}$. The sample was not gated and carrier density could be estimated at $n\sim10^{12}\;\mathrm{cm^{-2}}$ from previous measurements on similar samples. For comparison, we have included in the figure earlier data taken on a monolayer sample taken from Ref.\cite{Betz2013nphys} measured in an helium bath cryostat. The reference sample illustrates three of the four regimes introduced in Fig.\ref{Brunelfig1}: the Wiedemann-Franz, Bloch-Gr\"uneisen and the supercollision regimes. The BLG sample shows quite similar behavior, in particular the $T^4$-dependence from which we deduce the coupling constant $\Sigma=0.8\;\mathrm{mW/m^2/K^4}$. The value is very close to that reported of doped MLG \cite{Betz2012prl}; this similarity between phonon cooling in  MLGs and BLGs is not surprising in our experimental doping range  $\epsilon_F\sim 150\;\mathrm{meV}$ (see Ref.\cite{Viljas2010prb}). Our preliminary study needs however to be extended over a broad carrier doping range.

\begin{figure}[hh]
\includegraphics[width=7cm]{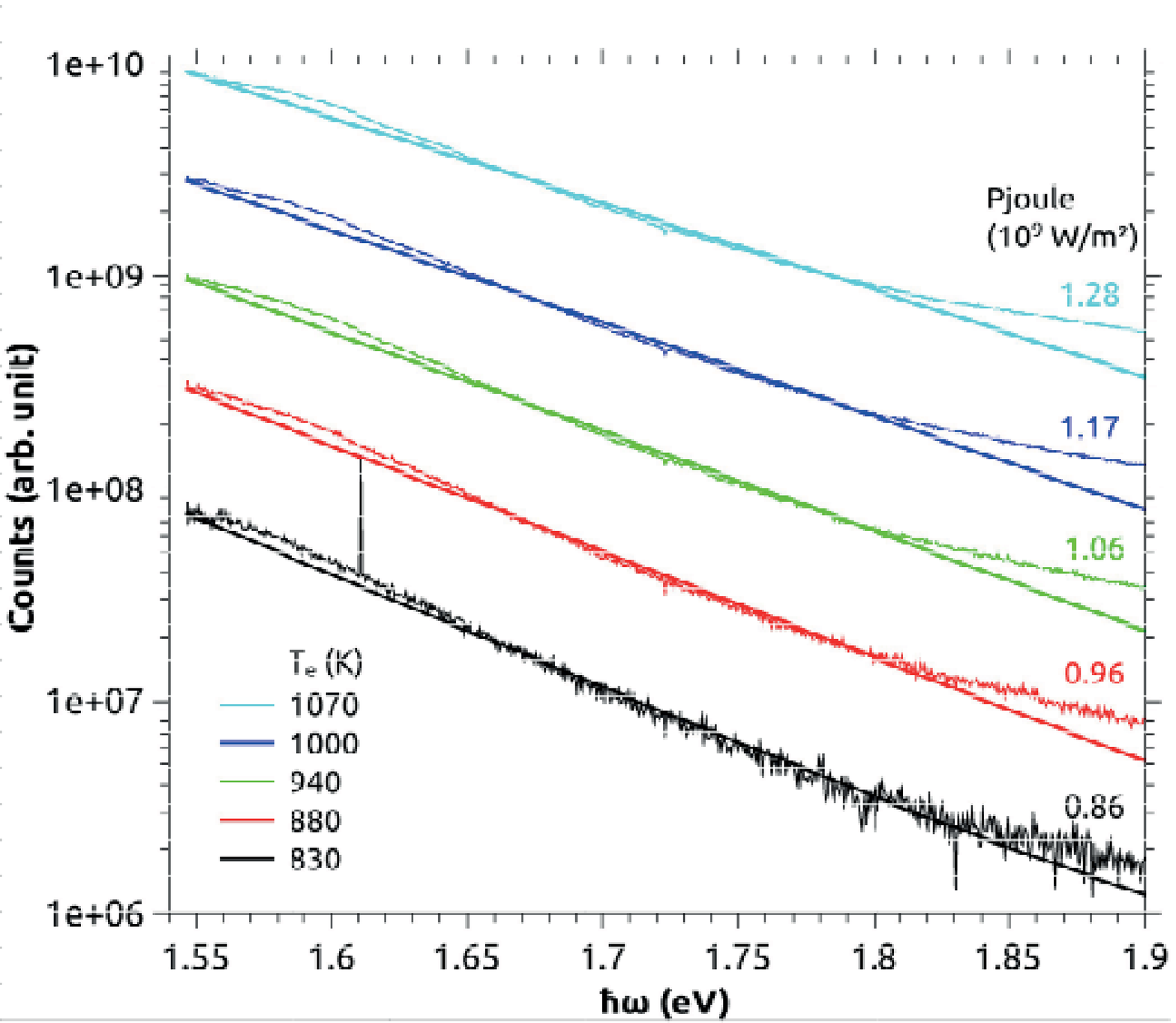}
\includegraphics[width=7cm]{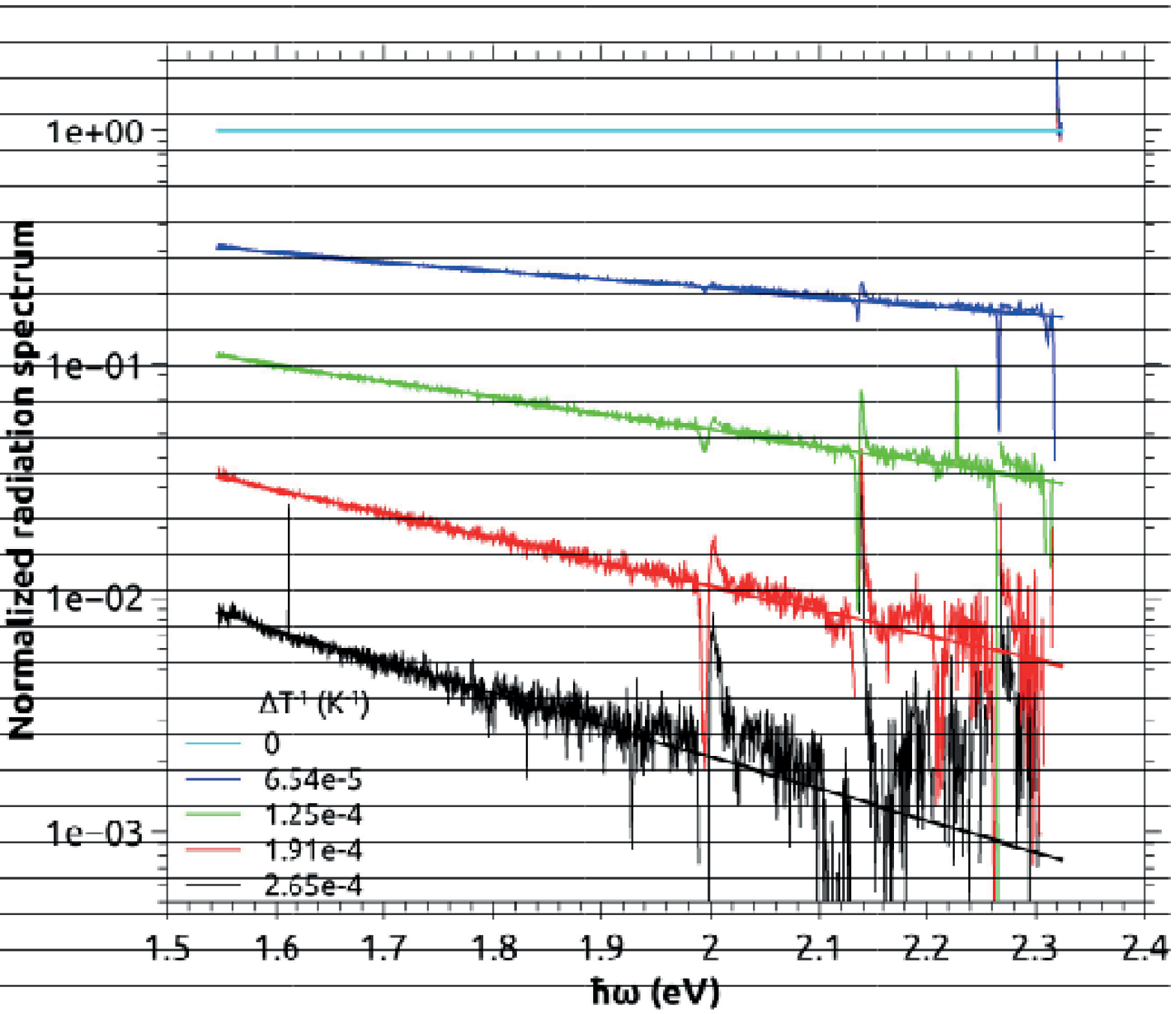}
\caption{Black-body radiation of the graphene channel for selected bias voltages. (a) Relative temperature measurements from an exponential fit of the spectra normalized to a reference spectrum recorded at 3.3~V. (b) Absolute temperature measurement from a fit of the spectra to the Planck law after correcting for the instrumental spectral response function.}\label{Brunelfig6}
\end{figure}

\begin{figure}[hh]
\includegraphics[width=9cm]{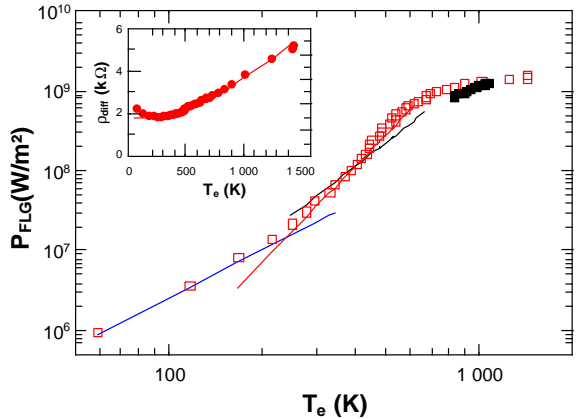}
\caption{Onset of optical phonon cooling measured by both noise thermometry (red squares) and black-body radiation (black squares). The sample, of dimensions $L\times W\simeq2\times4\;\mathrm{\mu m^2}$, is a few-layer graphene foil (FLG) exfoliated on silica; it is naturally doped. The derivation of the  electronic temperature from the black body spectrum  is detailed in Fig.\ref{Brunelfig6}. Blue, red and black solid lines are power law guidelines representing the three mechanisms of acoustic phonon cooling ($\delta=2$, $4$ and $3$ respectively see Figs.\ref{Brunelfig2} and \ref{Brunelfig5}). The optical phonon contribution shows up as a deviation from the $\delta=3$  power law above $500\;\mathrm{K}$. The inset shows the differential resistivity as function of electronic temperature. A phonon resistivity contribution, linear in temperature for  $T_e\gtrsim 400\;\mathrm{K}$, corresponds to  the hot phonon regime $T_{ph}\sim T_e/10\gtrsim 40\;\mathrm{K}$, which is a typical value for the Bloch-Gr\"uneisen temperature at intermediate doping $n_s\lesssim 10^{12}\;\mathrm{cm^{-2}}$.
   }\label{Brunelfig7}
\end{figure}

The interpretation of the Johnson-Nyquist microwave noise as a measurement of the black-body radiation of the electron gas through a single mode waveguide is well established, nevertheless our setup provides a unique way to measure this thermal radiation on the two extreme ends of the spectrum (namely the microwave and the visible ranges) for the very same electron gas. For large enough bias, this double measurement allows us to cross-check the calibrations of our electronic temperature measurements.
The black body radiation of hot electrons is collected by a long working distance microscope objective ($0.7$ numerical aperture) and dispersed in a $30\;\mathrm{cm}$ grating spectrometer. The signal is measured by a nitrogen cooled Si CCD camera. The sample is a few layer graphene flake of dimensions $L \times W=2\times 4\;\mathrm{\mu m^2}$ exfoliated on SiO$_2$. The carrier density is estimated at $n\simeq 10^{12}\;\mathrm{cm^{-2}}$ as previously for the BLG sample. The black-body radiation was measured in a second run where voltage cranked up from $V_{ds}=2.8\;\mathrm{V}$ but was interrupted at $3.3\;\mathrm{V}$ by sample breakdown. In order to estimate the absolute electronic temperature $T_e$, we fitted the spectra to the Planck law, after correcting the data from the spectral response function of the setup (Fig.\ref{Brunelfig6}-a). Ideally, this latter could be measured with a calibrated black-body. Here, we directly corrected the spectra from the objective and cryostat window
transmission and from the grating efficiency and CCD detector response functions. We also took into account signal modulations due to the interferences at the Si/SiO2/graphene interfaces behaving like a Fabry-P\'erot interferometer \cite{Yoon2009}. However, due to the limited spectral range in which this correction is applicable, the accuracy of the fits is limited to about $100\;\mathrm{K}$. As seen in Fig.\ref{Brunelfig7}, the electron temperature range ($800$-$1100\;\mathrm{K}$ for $P_J=0.85$-$1.3\times10^9\;\mathrm{W.m}^{-2}$) is consistent with the electron temperature deduced by the noise thermometry in this bias voltage range. To get rid of spectral correction issues, we also plotted (Fig\ref{Brunelfig6}-b) the black-body radiation spectra divided by the highest temperature spectrum. The very good linearity of the curves in a semi-log plot (as expected from the Planck law in the high frequency limit) gives an accurate relative temperature measurement scheme that is consistent with the corresponding noise thermometry data (Fig.\ref{Brunelfig7}). Using the same optical setup we also performed Raman spectroscopy to estimate the AC phonon bath temperature using the shift of the OP Raman peaks \cite{Calizo2007}. We found that the AC-phonon temperature remains one order of magnitude lower than that of electrons below $T_e=500\;\mathrm{K}$ in agreement with our estimate of Kapitza coupling \cite{Betz2012prl}. However at larger bias the phonon temperature clearly moves away from the base plate temperature approaching the electronic temperature for  $P\gtrsim10^9\;\mathrm{W.m}^{-2}$.

Figure \ref{Brunelfig7} shows the $P(T_e)$ data measured in the few-layer sample described above. In this  sample, we could reach very large voltage ($V_{ds}\lesssim 4\;\mathrm{V}$) and access electronic temperatures $T_e\lesssim 2000\;\mathrm{K}$. In particular, the temperature range exceeds the usual $T_e\lesssim 600\;\mathrm{K}$ window investigated in Fig.\ref{Brunelfig5}. The thermal energy $k_BT_e\lesssim0.15\;\mathrm{eV}$ approaching typical OP-phonons energies : $\hbar\Omega_\Gamma=0.20\;\mathrm{eV}$, $\hbar\Omega_K=0.16\;\mathrm{eV}$ for intrinsic graphene phonons. As for the bilayer of Fig.\ref{Brunelfig5}, the sample was not gated and presumably naturally p-doped. At a given measuring frequency, the accuracy of noise thermometry degrades at high bias due to the overwhelming contribution of the $1/f$-noise that scales $\propto I^2$ as compared to the sublinear dependence for the shot noise \cite{Betz2012prl}. As mentioned above, in this high temperature regime the spectral density being shifted toward infrared and optical bands, a similar resolution is expected by direct optical measurement of the black-body radiation (see Fig.\ref{Brunelfig1}-lower). The corresponding data are represented by black squares in the figure and superimpose onto the noise thermometry data. Remarkably these two independent determinations (shot noise and black body radiation) taken from the opposite ends of the emission spectrum ($10\;\mathrm{\mu eV}$ for RF and $1\;\mathrm{eV}$ for optics) give very consistent estimates of the electronic temperature.

We now discuss various cooling regimes observed in Fig.\ref{Brunelfig7}. As for the bilayer sample of Fig.\ref{Brunelfig5}, we retrieve the typical $T_e^2$ and $T_e^4$ dependencies for Wiedemann Franz and Bloch-Gr\"uneisen regimes  $T_e\lesssim400\;\mathrm{K}$ ($T_{ph}\lesssim40\;\mathrm{K}$). By contrast, $P(T_e)$ deviates from a power law above $500\;\mathrm{K}$ and takes the form of an exponential increase characteristic of the onset of an optical phonon contribution \cite{Viljas2010prb}. The departure from Bloch-Gr\"uneisen regime above $400\;\mathrm{K}$ is corroborated by the bias dependence of the differential resistivity plotted as function of the electronic temperature in the inset of Fig.\ref{Brunelfig7}. We observe the transition from a disorder limited constant resistivity at low temperature to a linear dependence at high temperature above $T_e\approx400\;\mathrm{K}$ ($T_{ph}\gtrsim40\;\mathrm{K}$). It is reminiscent of the onset of phonon resistivity reported in Ref.\cite{Efetov2010prl} and analyzed in Ref.\cite{Park2014nl} if one considers that in our out of equilibrium experiment the phonon temperature is given by $T_{ph}\sim T_e/10$ according to the ratio electron-phonon and phonon-substrate (Kapitza) coupling constants \cite{Betz2012prl}. The slope $d\rho/dT\simeq3\;\mathrm{Ohm/K}$  is however  larger than typical values $d\rho/dT\simeq1\;\mathrm{Ohm/K}$ for BLGs.  In our multilayer sample the fit to the $\Sigma T_e^4$ law gives $\Sigma\simeq4.5\;\mathrm{mW m^{-2}K^{-4}}$ which is typical of doped graphene where $T_{BG}\sim 40$--$60\;\mathrm{K}$. For completeness we have included in the figure a line $A T_e^3$ for supercollisions above $T_{BG}$ taking a prefactor $A=1.8\;\mathrm{W m^{-2} K^{-3}}$ which is also typical of doped graphene \cite{Betz2013nphys}. This shows in particular that supercollisions,  dominating AC-phonon cooling, cannot account for the observed cooling power above $400\;\mathrm{K}$. We are thus faced in Fig.\ref{Brunelfig7} with the onset of OP-phonon cooling in our few-layer graphene sample. This is corroborated by the observation of the increase in AC-phonon temperature observed in Raman spectroscopy. In the absence of specific theoretical prediction for few layer graphene, we cannot make a quantitative analysis of the data. However we can note that the order of magnitude $P\gtrsim 10^9\;\mathrm{W m^{-2}}$  at $1000\;\mathrm{K}$ compares favorably with recent observation in  BLG \cite{Laitinen2014arXiv}. At about $T_e=1000\;\mathrm{K}$ the increase of the phonon temperature leads to a saturation of the cooling power as seen in Fig.\ref{Brunelfig7}.

In total, we have shown the first set of measurements obtained with a new hybrid setup combining electrical and optical excitation and detection schemes for the study of hot carrier effects in few-layer graphene. The comparison of phonon cooling powers measured in this new setup with previous works allows us to validate the approach and to retrieve all the cooling regimes known for AC phonon coupling at low bias that are characterized by specific power laws. In addition, by cranking up the excitation power, we reached electronic temperatures above 1000~K. In this regime, noise thermometry measurements become more challenging but in contrast the accuracy of direct black-body optical detection takes over. Importantly, the cross-over regime allowed us to make sure that both thermometries give consistent results, validating the calibration protocols. Finally, this high-bias regime revealed the onset of new energy relaxation pathways implying either intrinsic or possibly remote optical phonon modes, that show up as a sudden increase in the cooling power above 500~K.

\begin{acknowledgments}
The research has been supported by the contracts ANR-2010-BLAN-MIGRAQUEL, and the EU Graphene flagship project (contract no. 604391). The work of Q.W. was supported by a DGA-MRIS scholarship. C.V. is a member of ``Institut Universitaire de France''. We thank P. Hakonen, P. Virtanen, T. Sohier and F. Mauri for fruitful discussions.
\end{acknowledgments}

\end{document}